\documentclass[]{interact}

\usepackage{epstopdf}
\usepackage[caption=false]{subfig}

\usepackage[numbers,sort&compress]{natbib}
\bibpunct[, ]{[}{]}{,}{n}{,}{,}

\usepackage{xcolor}
\usepackage{comment}

\renewcommand\bibfont{\fontsize{10}{12}\selectfont}
\makeatletter
\def\NAT@def@citea{\def\@citea{\NAT@separator}}
\makeatother

\theoremstyle{plain}
\newtheorem{theorem}{Theorem}[section]

\theoremstyle{definition}

\theoremstyle{remark}

\def\boxit#1{\vbox{\hrule\hbox{\vrule\kern6pt
\vbox{\kern6pt#1\kern6pt}\kern6pt\vrule}\hrule}}

\newcommand{\ps}[1]{{\color{black}#1}}

\begin{document}

\articletype{}

\title{Robust Variable Selection under Cellwise Contamination}

\author{
\name{Peng Su\textsuperscript{*,a}\thanks{*Corresponding Author: Peng Su, peng.su@sydney.edu.au.} Garth Tarr\textsuperscript{a} and Samuel Muller\textsuperscript{a,b}}
\affil{\textsuperscript{a}School of Mathematics and Statistics, The University of Sydney, NSW 2006, Australia;\textsuperscript{b}School of Mathematical and Physical Sciences, Macquarie University, NSW 2109, Australia}
}

\maketitle

\begin{abstract}
Cellwise outliers are widespread in real world data analysis. Traditional robust methods may fail when applied to datasets under such contamination. We introduce a variable selection procedure, that uses the Gaussian rank estimator to obtain an initial empirical covariance matrix among the response and potential predictors. We re-parameterise the classical linear regression model design matrix and the response vector such that we are able to take advantage of these robustly estimated components before applying the adaptive Lasso to obtain consistent variable selection results. 
The procedure is robust to cellwise outliers in low and high-dimensional settings. Empirical results show good performance compared with recently proposed robust techniques, particularly in the challenging environment when contamination rates are high but the magnitude of outliers is moderate.
\end{abstract}


\section{Introduction}
\label{intro}

Consider a linear regression model where $n$ observations are modelled through,
\begin{equation}
y_i =\bm{x}_i^\top\bm{\beta}^{} +\varepsilon_i,\quad i = 1,\ldots, n,
\label{modelfunc}
\end{equation}
where $y_i$ is the response, $\bm{x}_i\in\mathbb{R}^{p}$ is the vector of predictors, $\bm{\beta}^{}\in\mathbb{R}^{p}$ is the vector of regression parameters and $\varepsilon_i$ is an independent random error with variance $\sigma_\varepsilon^2$. \ps{We can write this more concisesly as ${\bm{y} =\bm{X}\bm{\beta}^{} +\bm{\varepsilon}}$, where $\bm{y} = (y_1,\ldots, y_n)^\top$, $\bm{X} = (\bm{x}_1,\ldots,\bm{x}_n)^\top$ and $\bm{\varepsilon} = (\varepsilon_1,\ldots,\varepsilon_n)^\top$.} If an intercept is included in model \eqref{modelfunc}, \ps{we alternatively write it as 
$$y_i = \beta_0^{} + \bm{x}_i^\top\bm{\beta}^{} +\varepsilon_i,$$ 
where $\beta_0^{}$ denotes the intercept parameter.}
For simplicity in notation, if not otherwise mentioned, we write regression models without an intercept term. 

\begin{figure}[!htp]
	\centering
	\includegraphics[width=10cm]{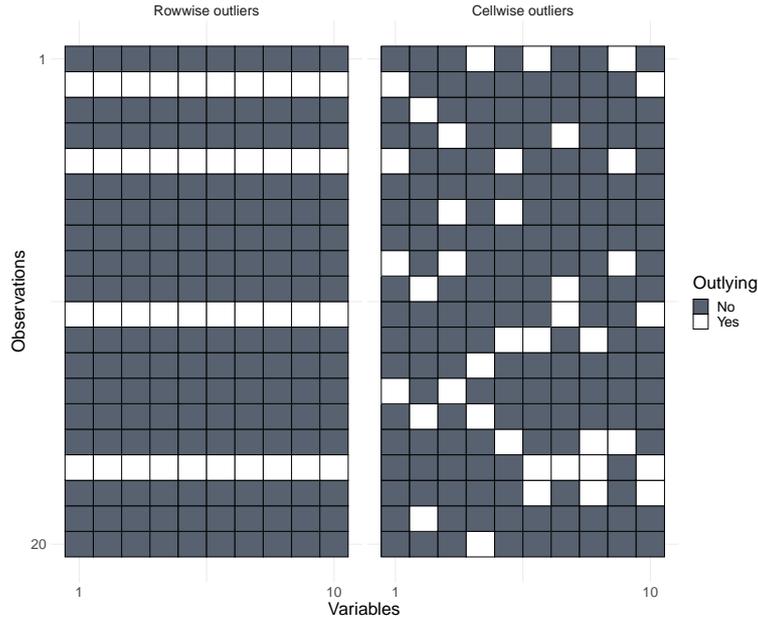}
	\caption{Rowwise and cellwise outliers: The outlying cells are rendered in white and the uncontaminated cells are shown in gray. For both panels, $20\%$ of the cells are contaminated. However, the left panel has 4 out of 20 rows outlying, while the right panel has 18 rows outlying.}
	\label{outliers}
\end{figure}

It is commonly believed that raw datasets contain about $1\% - 10\%$ outliers \cite{hampel1986robust}. Usually, the term outlier refers to rowwise outliers, as visualized in the left panel of Figure \ref{outliers}, where all $p$ elements of an entire row together `identify' the $i$-th row as outlying. 
Traditional robust regression estimators such as MM-estimators \cite{Yohai_1987} handle rowwise outliers by mitigating the influence of outlying rows through a robust loss function. The MM-estimator can be expressed as,
 \ps{$$
\hat{\bm{\beta}}_{\text{MM}} =\underset{{\bm{b}}}{\operatorname{argmin}}\sum_{i = 1}^{n}\rho\left(\frac{y_i -{\bm{x}_i^\top}\bm{b}}{\hat{\sigma}_{\varepsilon}}\right),
$$
where $\bm {b} = (b_1, \ldots, b_p)^\top$ indicates possible estimates of ${\bm{\beta}}$,
where $\rho(\cdot)$ is a robust loss function and $\hat{\sigma}_{\varepsilon}$ is an initial robust scale estimate of $\varepsilon$.} 
These robust estimators are less sensitive to rowwise outliers because the robust loss function $\rho(\cdot)$ downweights outlying rows as identified by their large residuals. 

Nevertheless, these estimators can cause overfitting and have poor robustness and efficiency properties when the $p$ versus $n$ ratio is high \cite{maronna_high_2015, smucler_highly_2015}. To solve this problem, 
Smucler and Yohai \cite{smucler2017robust} and Chang et al. \cite{chang_robust_2018} combine robust MM-estimator with the adaptive Lasso penalty (MM-ALasso),
$$
\hat{\bm{\beta}}_{\text{MM-ALasso}}=\underset{{\bm{b}}}{\operatorname{argmin}}\sum_{i=1}^n\rho\left(\frac{y_i -{\bm{x}_i^\top}\bm{b}}{\hat{\sigma}_{\varepsilon}}\right) +\lambda\sum_{j = 1}^{p}\omega_j |b_j|,
$$
where $\lambda$ is a tuning parameter and $\omega_j = 1/|\tilde\beta_j|, j = 1,\ldots, p$ are adaptive weights obtained from initial consistent estimates $\tilde\beta_j$. 
Other methods such as robust least angle regression (RLars) \cite{khan2007robust} and robust sparse least trimmed square (sLTS) \cite{alfons_sparse_2013} also perform well under rowwise contamination.

However, the rowwise outlier paradigm (visualized in the left panel of Figure \ref{outliers}) is often too narrow a framework. For some datasets, most cells in a row are clean, while just a few cells are contaminated. Alqallaf et al. \cite{alqallaf_propagation_2009} describe the propagation of cellwise outliers under the independent contamination model. For a contamination rate $e$ of cells, the expected proportion of contaminated observation rows is $1-(1-e)^p$, which on average rapidly exceeds $50\%$ with increasing dimension $p$, and this can even be more rapid in special cases, as shown in the right panel of Figure \ref{outliers}.

Figure \ref{outliers} shows two scenarios with the same proportion of contaminated cells in the design matrix, i.e., 20\% of entries are contaminated overall. However, the scenario in the left panel has only a few observation rows contaminated, whereas 90\% of rows in the right panel have at least one contaminated cell. Traditional robust statistical methods are characterised by their breakdown point, which is the proportion of rows that can be contaminated before the method fails. The most robust methods from the rowwise paradigm have a breakdown point of 50\%. It is clear that rowwise robust methods provide limited protection against cellwise outliers because even a small overall proportion of contaminated cells may result in most rows having at least one outlying cell. That is, the proportion of rowwise contamination can very quickly exceed 50\%, rendering even the most robust traditional methods ineffective.

A natural choice to handle cellwise outliers is to perform outlier detection first. For example, Rousseeuw and Bossch \cite{rousseeuw_detecting_2018} predict the value of a cell and then flag a cell as outlying if its observed value and its predicted value differ by too much.
Debruyne et al. \cite{debruyne_outlyingness_2019} regard all outliers as rowwise outliers and then detect which cells contribute more to the outlying rows.
Raymaekers and Rousseeuw \cite{raymaekers_flagging_2019} detect outlying cells for each row by proposing a `cellHandler' technique, which combines a Lasso penalty \cite{tibshirani_regression_1996} with a stepwise application of constructed cutoff values. 
After an initial outlier detection, there are two natural choices for the next step: (1) impute non-outlying values for those cells flagged as outlying, as done in \cite{rousseeuw_detecting_2018,debruyne_outlyingness_2019}; (2) regard detected outliers as missing and then obtain robust estimation directly, as in \cite{raymaekers_flagging_2019}.

Very little work has been done in the context of robust regression under cellwise contamination. 
{\"O}llerer et al. \cite{ollerer_shooting_2016} propose the Shooting S algorithm to perform robust regression. First, a robust imputation of the design matrix is obtained. Then the Shooting S algorithm is used to obtain a robust estimation of regression coefficients. Bottmer et al. \cite{bottmer_sparse_2021} combine the Shooting S algorithm with the Lasso to obtain sparse regression results. 
Leung et al. \cite{leung_robust_2016} propose a three-step regression method. First, cellwise outliers are detected marginally; second, robust multivariate location and scatter estimators are applied to obtain robust location and covariance estimates; finally, the regression coefficients are computed from the estimates obtained in the second step. 
Filzmoser et al. \cite{filzmoser_cellwise_2020} propose a cellwise robust M-estimator by replacing the detected outliers with imputed values. 

To obtain robust variable selection results under cellwise contamination, especially in high-dimensional cases, we propose GR-ALasso, a procedure to perform variable selection based on a robust estimate of the covariance matrix among response and predictors. 
In Section~\ref{meth}, we use a cellwise robust estimator to obtain an initial empirical covariance matrix. Then we incorporate the estimated covariance matrix in a re-parameterised linear model to achieve robust variable selection. Empirical works show that our proposed methods are robust to cellwise outliers in comparison with recently proposed approaches and traditional rowwise robust techniques. Simulation results are presented in Section~\ref{simu} and real data applications in Section~\ref{data} demonstrate the utility of our proposed method. R functions are available on the GitHub page of the first author (https://github.com/PengSU517/robcovsel). 

\section{Variable selection based on robust covariances}
\label{meth}
\subsection{Robust estimation of the covariance matrix}
 \ps{We begin by introducing key notations and definitions. Let $\bm{Z} = (\bm{y},\bm{X})\in\mathbb{R}^{n\times(p+1)}$ as the data matrix, where the response vector $\bm{y}$ and the design matrix $\bm{X}$ follow from the definition of the regression model \eqref{modelfunc}. We define $\bm{z}^j$ as the $j$-th column of $\bm{Z}$. The covariance matrix of $\bm{Z}$ is denoted as $\bm{\Sigma}$ with $\bm{R}$ denoting its corresponding correlation matrix. Additionally, $\bm{S}$ denotes the diagonal matrix, which contains the scale estimates of each variable, that is
$\bm{S} =\operatorname{Diag}(\sigma_{z^1},\ldots,\sigma_{z^{p+1}})$ where $\sigma_{z^j}$ indicates the scale parameter of variable $\bm{z}^j$. The corresponding estimates of these matrices are denoted as $\hat{\bm{\Sigma}}$, $\hat{\bm{R}}$, and $\hat{\bm{S}}$, respectively.}

Given a pair of columns $\bm{z}^j$ and $\bm{z}^k$, 
an efficient robust correlation estimator is the Gaussian rank (GR) correlation \cite{boudt_gaussian_2012} defined as the sample correlation estimated from the normal scores of the data. For $z_{ij}$, the $i$-th observation of the $j$-th variable, we compute,
\[
	\tilde z_{ij} =\Phi^{- 1 }\left(\frac{\operatorname{Rank} ( z _{i j } ) }{n + 1 }\right),
\]
where $\operatorname{Rank} ( z_{ij} )$ denotes the rank of $z_{ij}$ among all $n$ elements of $\bm{z}^j$, $\Phi(\cdot)$ represents the univariate standard normal cumulative distribution function and $\Phi^{-1}(\cdot)$ is the corresponding quantile function. After obtaining a pseudo dataset $\tilde{\bm Z} = (\tilde z_{ij})_{n\times (p+1)}$, we calculate the GR correlation matrix by computing the Pearson correlation matrix of $\tilde{\bm{Z}}$. Compared with other nonparametric correlation estimators, the GR correlation shows good robustness \cite{boudt_gaussian_2012}, consistency and considerable efficiency \cite{amengual2022gaussian}. Due to its excellent properties, it is considered as a plug-in estimator for precision matrices \cite{ollerer_robust_2015}. Another crucial advantage of the GR correlation is that the estimated correlation matrix is necessarily positive semi-definite even in high-dimensional cases. This ensures that the regression optimization process for \eqref{lossfunc} is completely convex.

As alternatives, several other nonparametric correlations can be employed to obtain robust correlation matrices. These include the Quadrant correlation \cite{blomqvist1950measure}, Spearman correlation \cite{spearman1904proof}, and pairwise correlations like the Gnanadesikan-Kettenring correlation (GK) \cite{gnanadesikan_robust_1972}. These methods are extensively studied for their robustness and reliability in various applications \cite{tarr_robust_2016, loh_rob_precision_2018}. Researchers and practitioners may consider these approaches as viable options based on the specific characteristics and requirements of their datasets and analysis goals. In empirical work not shown here, we find that the pairwise correlations are not competitive with the GR estimator.
Robust scale estimators such as the median absolute deviation, $Q_n$ estimator \cite{rousseeuw_alternatives_1993} and $P_n$ estimator \cite{tarr_robust_2012} can be used to obtain robust \ps{scale (equivalent to standard deviation under normal distribution)} estimates. We advocate using the $Q_n$ estimator because of its robustness and efficiency.
After obtaining estimates of the scale parameters and the correlation matrix, we obtain a robust empirical covariance matrix $\hat{\bm{\Sigma }} =\hat{\bm{S}}\hat{\bm{R}}\hat{\bm{S}} $. 
In computations, we can also use $\hat{\bm{R}}$ directly then the scaling of predictors can be dealt with initially.


\subsection{Modification of the regression loss}
After obtaining an estimated covariance matrix $\hat{\bm{\Sigma}}$ from the GR estimator, we demonstrate how this can be used for variable selection. Loh and Wainwright \cite{loh_high-dimensional_2012} show that the objective loss function of the linear regression model \eqref{modelfunc} can be expressed as follows,
\begin{equation}
    \hat{\bm{\beta}}_{\text{GR}} =\underset{\bm{b}}{\operatorname{argmin}}\left\{n{\hat\Sigma}_{yy} + n\bm{b}^\top\hat{\bm{\Sigma}}_{\bm{x}\bm{x}}\bm{b} - 2n\bm{b}^\top\hat{\bm{\Sigma}}_{\bm{x}y}\right\},
	\label{lossfunc}
\end{equation}
where $\hat{{\Sigma}}_{yy}$ represents the estimated variance of $y$, $\hat{\bm{\Sigma}}_{\bm{x}y}$ represents the estimated covariance vector between $\bm{x}$ and $y$, $\hat{\bm{\Sigma}}_{\bm{x}\bm{x}}$ represents the estimated covariance matrix of $\bm{x}$. They are the components of the estimated covariance matrix $\hat{\bm{\Sigma}}$ among predictors and the response,
\[
	\hat{\bm{\Sigma}} =\begin{pmatrix}
		{\hat\Sigma}_{yy} &\hat{\bm{\Sigma}}_{\bm{x}y}^\top\\
		\hat{\bm{\Sigma}}_{\bm{x}y} &\hat{\bm{\Sigma}}_{\bm{x}\bm{x}}
	\end{pmatrix}.
\]
The solution of \eqref{lossfunc} can be expressed as $\hat{\bm{\beta}}_{\text{GR}} =\hat{\bm{\Sigma}}_{\bm{x}\bm{x}}^{-1}\hat{\bm{\Sigma}}_{\bm{x}y}$. Leung et al. \cite{leung_robust_2016} highlight that one way to obtain robust regression results under cellwise contamination is through using a robust counterpart of an empirical covariance matrix.

The objective loss \eqref{lossfunc} can be more elegantly expressed. Given that $\hat{\bm{\Sigma}}$ is positive semi-definite, let the square root of $\hat{\bm{\Sigma}}$ be $ \hat{\bm{\Sigma}}^{1/2} = ({\bm{v}},{\bm{W}})$ where ${\bm{v}}$ is the first column of $\hat{\bm{\Sigma}}^{1/2}$ and ${\bm{W}}$ is a matrix consisting of the remaining columns of $\hat{\bm{\Sigma}}^{1/2}$. 
 \ps{Taking note of the following relationships, ${\bm{v}}^\top{\bm{v}} ={\hat{\Sigma}}_{yy}$, ${\bm{W}}^\top{\bm{v}} =\hat{\bm{\Sigma}}_{\bm{x}y}$ and ${\bm{W}}^\top{\bm{W}} =\hat{\bm{\Sigma}}_{\bm{x}\bm{x}},$ we rewrite the objective loss \eqref{lossfunc} as follows,}
\begin{equation}
	\hat{\bm{\beta}} =\underset{\bm{b}}{\operatorname{argmin}}\left\{n\|{\bm{v}} -{\bm{W}}\bm{b}\|_2^2\right\},
	\label{lossfunc2}
\end{equation}
which presents a classic quadratic optimization problem.

To achieve sparse variable selection results, we propose an adaptive Lasso estimator based on GR correlations (GR-ALasso), resulting in a regularized objective loss function,
\begin{equation}
	\hat{\bm{\beta}}_\mathrm{GR-ALasso} =\underset{\bm{b}}{\operatorname{argmin}}\left\{n\|{\bm{v}} -{\bm{W}}\bm{b}\|_2^2 +\lambda\sum_{j = 1}^{p}\omega_j |{b}_j|\right\}
	\label{aLassoloss},
\end{equation}
where $\lambda$ is the tuning parameter, ${\omega}_j = 1/{\tilde{\beta}_j}$ and
${\tilde{\beta}}_j$ is an initial consistent estimate of ${\beta}_j$. When $p\ll n$, we compute $\tilde{\bm{\beta}} = (\tilde{\beta}_1,\ldots,\tilde{\beta}_p)^\top =\hat{\bm{\Sigma}}_{\bm{x}\bm{x}}^{-1}\hat{\bm{\Sigma}}_{\bm{x}y}$ as an initial consistent estimator. \ps{In high-dimensional cases, ridge regression can be used to obtain an initial estimate, $\tilde{\bm{\beta}} =\left(\hat{\bm{\Sigma}}_{\bm{x}\bm{x}} +\kappa\right)^{-1}\hat{\bm{\Sigma}}_{\bm{x}y}$, where $\kappa$ is a tuning parameter.} If an intercept term is included, the intercept can be estimated by 
$\hat{\beta}_0 =\hat{\mu}_y -\hat{\bm{\mu}}_{\bm{x}} ^\top\hat{\bm\beta}$, where $\hat\mu_y$ indicates the robust location estimate of the response and $\hat{\bm\mu}_{\bm{x}}$ indicates the robust location estimate of the predictors. 

\subsection{Selection of the tuning parameter}

The optimal value for the tuning parameter $\lambda$ in Lasso-type problems can be determined by calculating solutions for a range of $\lambda$ values and then choosing the value that results in the smallest cross-validation (CV) error, as described in \cite{hastie2009elements}. Further, in order to have a more parsimonious model, we use the ``one-standard-error'' rule \cite{hastie2009elements}, which selects the most parsimonious model that lies within one standard error of the smallest CV error.

However, CV with the original dataset $\bm{Z}$ is not appropriate under cellwise contamination. Due to the propagation of cellwise outliers, the row contamination rate can easily exceed $50\%$, which potentially means most prediction errors obtained from CV are outlying. In this instance, we recommend running CV with the pseudo-dataset $\tilde{\bm Z}$. In the sections below, we show the empirical results of our proposed methods using the 5-fold CV approach, along with the ``one-standard-error'' rule, applied to the pseudo-dataset.

\subsection{Theoretical properties}
\label{theorem}

The GR estimator and GR-ALasso demonstrate strong theoretical properties, including good breakdown properties, which can be inferred directly from previous works \cite{rousseeuw_alternatives_1993, boudt_gaussian_2012}. Moreover, we establish the oracle property of the GR-ALasso following the approach introduced by \cite{zou_adaptive_2006}.

 \ps{

We use the following symbols and definitions to establish our main theoretical result. For ease of reading, we write $\hat{\bm{\beta}}$ for the GR-ALasso estimator in this subsection and corresponding appendix.
For the $j$-th variable $z^j$ in the random vector $\bm{z} = (y,\bm{x}^{\top})^{\top}$, we denote its marginal cumulative distribution function (CDF) as $F_j(\cdot)$. Additionally, we define a random vector $\check{\bm{z}} = (\check{z}^1,\check{z}^2,\ldots,\check{z}^{p+1})^{\top}$, where $\check{z}^j =\Phi^{-1}(F_j(z^j))$ and $\Phi$ indicates the CDF of the standard normal distribution. This means $\check{\bm{z}}$ is a standardized version of $\bm{z}$ if $\bm{z}$ follows a multivariate normal distribution.}
Next, we define $\bm{\rho} =\mathrm{vecl}(\bm{R})$, where $\mathrm{vecl}(\cdot)$ is a vector-type operator that stacks, by columns, the components in the strict lower triangle of a matrix, followed by an estimation $\hat{\bm{\rho}} =\mathrm{vecl}(\hat{\bm{R}}).$

\ps{
We define $\mathcal{I}(\bm{\rho})$ as the Fisher information matrix of $\bm{\rho}$ corresponding to $\check{\bm{z}}$ and $\bm{\nabla}\bm{\beta}^{}$ as the gradient of $\bm{\beta}^{}$ with respect to $\bm{\rho}^\top$, where more detailed structure of $\mathcal{I}(\bm{\rho})$ and $\bm{\nabla}\bm{\beta}^{}$ can be found in the appendix and in \cite{amengual2022gaussian}.
Let $\mathcal{A} =\{1,2,\ldots, p_0\}$ represent the set of active predictors, assuming without loss of generality that the first $p_0$ predictors are active. Then $\hat{\mathcal{A}}$ represents the set of selected variables corresponding to ${\hat{\bm{\beta}}}$, and ${\bm{\beta}}_{\mathcal{A}}$ represents the subvector of ${\bm{\beta}}$ that includes only the components corresponding to $\mathcal{A}$. 
}
For the subsequent analysis, we establish the following assumptions:

\begin{itemize}
\item A1: \ps{The random vector $\bm{z}$ follows a multivariate normal distribution $N(\bm{0},\bm{\Sigma})$.}
\item A2: The penalty $\lambda/\sqrt{n}\rightarrow 0$ and $\lambda\rightarrow\infty$ when $n\rightarrow\infty$.
\end{itemize}
 \ps{When the random vector $\bm{z}$ follows a multivariate normal distribution, each marginal variable $z^j$ follows a univariate normal distribution. Then we can obtain the consistent $Q_n$ scale estimator \cite{rousseeuw_alternatives_1993} for each $z^j$ since the distribution type is known.} The Gaussian rank estimator is fully efficient under multivariate normal distributions and remains consistent for other distributions \cite{amengual2022gaussian}. A2 is a common assumption for adaptive Lasso type estimators which guarantees that the estimator is consistent when other assumptions hold \cite{zou_adaptive_2006}. 

\begin{theorem}
If assumptions A1-A2 hold, the GR-ALasso estimator satisfies the following:
\begin{enumerate}
\item Asymptotic normality: 
 \ps{
\[
\sqrt{n}\left({\hat{\bm{\beta}}}_{\mathcal{A}}-{\bm{\beta}}_{\mathcal{A}}^{}\right)\stackrel{d}{\rightarrow} N\left(\bm{0},\bm{\nabla}\bm{\beta}_{\mathcal{A}}^{}\mathcal{I}^{-1}\left(\bm{\rho}\right)\bm{\nabla}\bm{\beta}_{\mathcal{A}}^{ \top}\right).
\]
}
\item Consistency in variable selection: $P(\hat{\mathcal{A}} ={\mathcal{A}})\rightarrow 1.$
\end{enumerate}
\label{theorem1}
\end{theorem}
The proof of \ref{theorem1} is shown in the appendix.
\section{Simulation studies}
\label{simu}

To evaluate the performance of GR-ALasso, we conduct a comparison of its performance with several other popular methods including Lasso \cite{tibshirani_regression_1996}, adaptive Lasso (ALasso) \cite{zou_adaptive_2006}, adaptive Lasso regularized MM-estimation (MM-ALasso) \cite{smucler2017robust}, robust least angle regression (RLars) \cite{khan2007robust}, sparse least trimmed squares (sLTS) \cite{alfons_sparse_2013} and sparse shooting S-estimator (sShootingS) \cite{bottmer_sparse_2021}.

\subsection{Low-dimensional settings: $p < n$}
\label{simu1}

Recall the linear regression model \eqref{modelfunc}. In our simulations, we set $n = 100$, $p = 20$, $\bm{\beta}^{} = (\bm{1}_5^\top,\bm 0_{15}^\top )^\top$, $\bm{x}_i$ is sampled from $N(\bm 0,\bm{\Sigma_{\bm{x}\bm{x}}})$ and $\varepsilon_i$ is sampled from $N(0,1^2)$.
The correlation structure among predictors is given by $\Sigma _{ij} = 0.5^{|i-j|}$, where $\Sigma _{ij}$ represents the correlation between the $i$-th and the $j$-th predictor. 
Contamination proportions are set as $ 2\%$, $5\%$ and $10\%$ for all predictors separately. Outlying cells of all predictors are randomly generated from $N(\gamma, 1)$ or $N(-\gamma, 1)$ with equal probability. We vary $\gamma$ over the set $\{2,4,6,8,10\}$ to simulate outliers with different magnitudes. 

\begin{figure}[!htp]
	\centering
	\includegraphics[width=14cm]{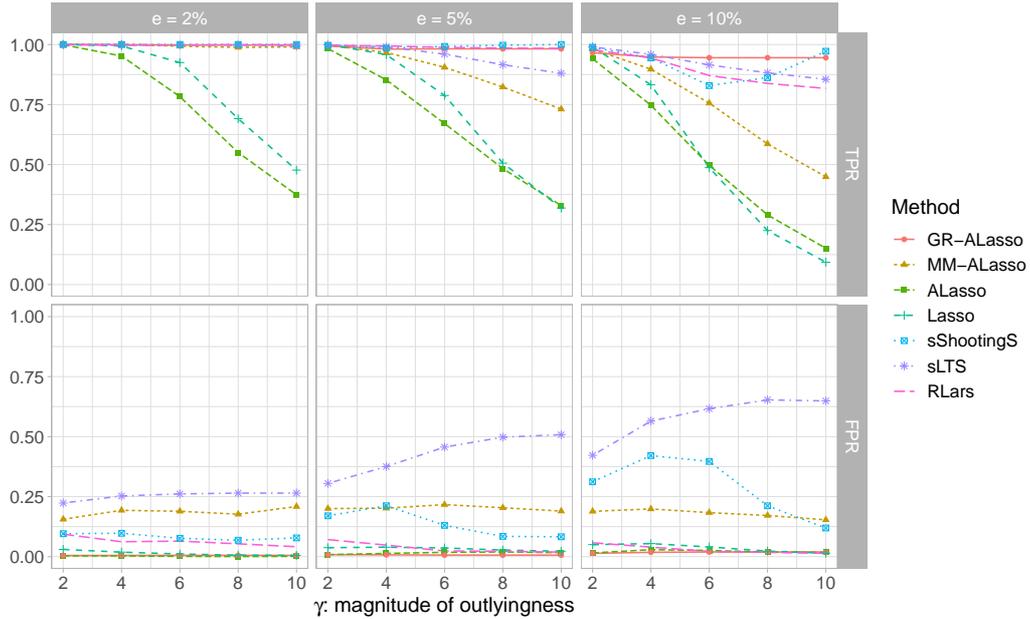}
	\caption{Selection results as summarised by the averages of TPRs and FPRs over 200 simulation runs when $p = 20$. Each column represents a contamination rate. The horizontal axis represents the magnitude of outlyingness and the vertical axis gives the TPRs and FPRs.}
	\label{fig20}
\end{figure}

The experiment is repeated $200$ times for each scenario. In each run, the whole dataset including the design matrix is regenerated. We use the true positive rate (TPR) and the false positive rate (FPR) to assess the variable selection performance of each method.
For both TPR and FPR, we report the average values from 200 simulations in Figure \ref{fig20}.

While prediction is not the main focus of this paper, for the sake of completeness, we also report in the appendix material prediction results by computing the averages of the mean squared prediction error (MSPE) from independent datasets, where $\mathrm{MSPE} = n^{-1}\sum_{i = 1}^n(\hat y_i - y_i)^2.$ We also use the mean squared error (MSE) of $\hat{\bm{\beta}}$ to assess the estimation results, where $\mathrm{MSE} = p^{-1}\sum_{j = 1}^p(\hat{\beta}_{j} -{\beta}_{j}^{})^2$. 

Figure \ref{fig20} reports the averages of TPRs and FPRs for the data generated using $p = 20$. 
With a $2\%$ contamination rate (as shown in the left column), most methods correctly identify all active predictors, except for Lasso and ALasso, which are not designed to handle outliers. Among other methods, GR-ALasso, RLars, and sShootingS tend to select fewer inactive predictors, demonstrating their superior selection abilities. These three methods are also the only three cellwise robust methods in comparison. On the other hand, the sLTS and MM-ALasso tend to select more inactive predictors, resulting in a higher probability of overfitting.
When the contamination rate increases to $5\%$ (as shown in the middle column), GR-ALasso and RLars still perform well. Meanwhile, the performance of sLTS and MM-ALasso deteriorates as their TPRs decrease with an increase in $\gamma$. The sShootingS also tends to select more inactive predictors in this scenario. Other methods show a decline in performance compared to the $2\%$ contamination scenario.
With a $10\%$ contamination rate (as shown in the right column), all methods show inferior performance. Among them, GR-ALasso has the best TPR (about $95\%$), although it is not as good as the low contamination scenarios. The TPR of RLars is only about $80\%$ when $\gamma = 10$. The sShootingS also demonstrates a lower TPR with moderate $\gamma$ values.

Overall, GR-ALasso outperforms other methods for all combinations of contamination proportion and outlier magnitude. Rowwise methods such as MM-ALasso and sLTS perform well with low contamination rates ($2\%$). Meanwhile, cellwise robust methods (RLars, sShootingS, GR-ALasso) work well with moderate contamination rates ($5\%$) but are still not effective enough with high contamination rates ($10\%$).

\subsection{High-dimensional settings: $p > n$}
\label{simu2}
In this section, we show the empirical findings in high-dimensional cases. We continue with the simulation settings from the second part of the previous section but now increase $p$ to be $200$ while keeping $n$ at $100$. We employ ridge estimates to calculate the adaptive weights for ALasso. For MM-ALasso, we calculate the adaptive weights through the use of MM-Ridge \cite{maronna2011robust}. for GR-ALasso, we compute adaptive weights with a ridge regularized objective loss \eqref{lossfunc2}. Results of TPR and FPR are reported in Figure \ref{fig200} and additional results are shown in the appendix. 

\begin{figure}[!htp]
	\centering
	\includegraphics[width=14cm]{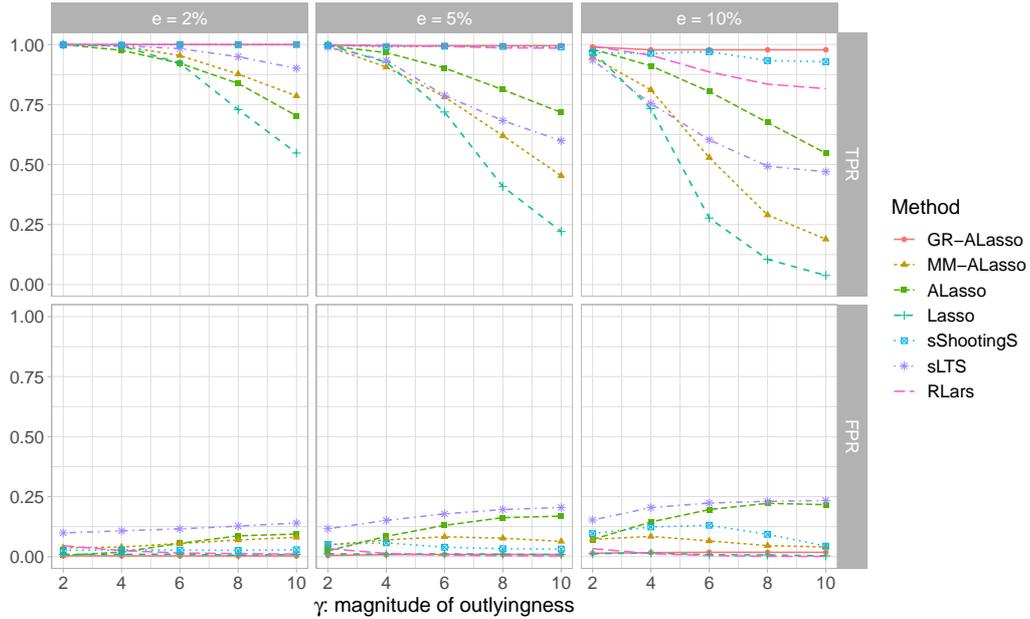}
	\caption{Selection results as summarised by the averages of TPRs and FPRs over 200 simulation runs for various contamination rates when $p = 200$. Each column represents a contamination rate. The horizontal axis represents the magnitude of outlyingness and the vertical axis gives the TPRs and FPRs.}
	\label{fig200}
\end{figure}

Figure \ref{fig200} shows simulation results under high-dimensional settings. In this more challenging scenario, the overall performance of GR-ALasso remains strong. 

With a low contamination rate (as shown in the left column of Figure \ref{fig200}), GR-ALasso and Rlars show outstanding performance, implying that a few outliers do not dramatically affect the variable selection accuracy in high-dimensional cases. The sShootingS shows slightly inferior FPR. On the other hand, MM-ALasso and sLTS demonstrate lower TPRs, indicating their ineffectiveness in high-dimensional cases even with only $2\%$ contamination. As anticipated, ALasso and Lasso still show poor performance. 

As the contamination rate rises to $5\%$ (as shown in the middle column of Figure \ref{fig200}), GR-ALasso, sShootingS, and Rlars maintain excellent performance, while other methods show declined performance compared to the low contamination case.

In the presence of a high contamination rate (as shown in the right column of Figure \ref{fig200}), all methods experience a decline in performance. Rlars is only able to select $80\%$ active predictors with a high magnitude of outliers ($\gamma = 10$). The sShootingS also shows a poor TPR when $\gamma = 10$ and shows a poor FPR when $\gamma = 6$. Balancing TPR and FPR becomes a challenge for the sShootingS in this scenario.

In conclusion, while other methods may not be as performant as in low-dimensional settings, GR-ALasso still demonstrates remarkable results even when faced with high magnitudes of contamination. As the GR-ALasso is specifically designed for variable selection, it is recommended to perform robust post-regression after model selection to improve prediction accuracy.

\section{Real data applications}
\label{data}

We show the performance of the proposed methods using two real datasets: the Boston housing data and a breast cancer data set. They are representative of applications in low and high-dimensional settings, respectively. 

\subsection{Boston housing data}
\label{boston}

We illustrate the effect of cellwise outlier propagation on the previously considered estimators using the Boston Housing dataset. This data is available at the UCI machine learning repository (https://archive.ics.uci.edu/ml/index.php) and is collected from $506$ census samples on $14$ different variables. The original objective of the study is to analyze the association between the median housing values (\textit{medv}) in Boston and the residents' willingness to pay for clean air, as well as the association between\textit{medv} and other variables in the dataset.

In our research, we only consider the $9$ quantitative variables that are extensively studied. 
First, we standardize all predictors with robust estimators of location and scale; here we use the median and $Q_n$. Then a sparse regression model is fitted for variable selection:
\[
	\begin{aligned}
		\log(medv) = & \ \beta_0^{} +\beta_1^{}\log(dis) +\beta_2^{} black +\beta_3^{} rm +\beta_4^{} ptratio\\ 
  & +\beta_5^{}\log(lstat) +\beta_6^{} age +\beta_7^{} tax +\beta_8^{} nox +\beta_9^{} \log(crim) +\varepsilon.\\
	\end{aligned}
\]
We compare the performance of 7 methods: Lasso \cite{tibshirani_regression_1996}, ALasso \cite{zou_adaptive_2006}, MM-ALasso \cite{smucler2017robust}, RLars \cite{khan2007robust}, sLTS \cite{alfons_sparse_2013} and sShootingS \cite{bottmer_sparse_2021}, as well as our newly proposed method GR-ALasso.

To measure the performance of variable selection, we generate $10$ additional random variables as known redundant predictors. These additional variables are generated from a zero-mean multivariate normal distribution with correlation structure $\Sigma_{ij} = 0.5^{|i-j|}$. Therefore, we now have 19 predictors to choose from. 

For each predictor, $5\%${of the} cells are replaced by cellwise outliers which are randomly generated from $N(10, 1)$ or $N(-10, 1)$ with equal probability. As a comparison, we also run simulations without any contamination to investigate how stable the various methods are when known outliers are present in the data. We repeat this process 200 times and then compute the selection rates of each variable (shown in Table \ref{app1}). 

\begin{table}[!htp]
	\centering
	\caption{Variable selection rates for Boston Housing data over 200 simulation runs. FPR represents the average false positive selection rate across all the ten redundant predictors and $e$ represents the contamination rate.}
	
	\resizebox{14cm}{!}{
\begin{tabular}{lllllllllllll}
\toprule
method & $e$ & $\log(lstat)$ & $rm$ & $\log(dis)$ & $tax$ & $ptratio$ & $nox$ & $age$ & $black$ & $\log(crim)$ & FPR\\ 
\midrule
GR-ALasso & 0 & 1 & 1 & 0.26 & 1 & 1 & 0 & 0 & 0 & 0 & 0\\ 
 & 0.05 & 1 & 1 & 0 & 0.96 & 1 & 0.03 & 0.03 & 0 & 0.06 & 0\\ 
 MM-ALasso & 0 & 1 & 1 & 0.99 & 1 & 1 & 0.98 & 1 & 1 & 0.84 & 0.84\\ 
 & 0.05 & 0.94 & 1 & 0.23 & 0.81 & 0.87 & 0.28 & 0.44 & 1 & 0.32 & 0.23\\ 
 ALasso & 0 & 1 & 1 & 0 & 0.98 & 1 & 0 & 0 & 1 & 0 & 0\\ 
 & 0.05 & 1 & 0.99 & 0.04 & 1 & 0.70 & 0.28 & 0.28 & 1 & 0.41 & 0\\ 
 Lasso & 0 & 1 & 1 & 0.01 & 1 & 1 & 0 & 0 & 1 & 0 & 0.01\\ 
 & 0.05 & 1 & 1 & 0.20 & 1 & 0.92 & 0.70 & 0.69 & 1 & 0.76 & 0.01\\ 
 sShootingS & 0 & 1 & 0 & 0 & 0 & 0 & 0 & 0 & 0 & 0 & 0\\ 
 & 0.05 & 1 & 0 & 0 & 0 & 0 & 0 & 0 & 0 & 0 & 0\\ 
 sLTS & 0 & 1 & 1 & 0 & 1 & 1 & 0 & 1 & 1 & 0 & 0.02\\ 
 & 0.05 & 0.96 & 0.94 & 0.16 & 0.96 & 0.99 & 0.30 & 0.48 & 1 & 0.36 & 0.11\\ 
 RLars & 0 & 1 & 1 & 1 & 1 & 1 & 0.86 & 0.13 & 1 & 0.86 & 0.33\\ 
 & 0.05 & 1 & 1 & 0.45 & 0.96 & 0.96 & 0.34 & 0.06 & 0.86 & 0.47 & 0.23\\ 
\bottomrule
\end{tabular}
	}
	\label{app1}%
\end{table}%

For all seven methods reported in Table \ref{app1}, we observe some unstable selection results between the contaminated and uncontaminated settings. GR-ALasso shows very consistent selection results for the two scenarios considered (without cellwise outliers and with 5\% cellwise outliers). 
MM-ALasso and sLTS select a large number of redundant predictors and show inconsistency for many original predictors. ALasso and Lasso also show some inconsistency even with good FPRs. On the other hand, the sShootingS stably selects extremely sparse models, almost exclusively selecting $\log(lstat)$ for both the uncontaminated and contaminated settings. 
RLars also shows inconsistent selection results for many predictors. These results are consistent with simulation results in low-dimensional settings.

\subsection{Breast cancer data}
\label{breast}

In the breast cancer dataset \cite{van_t_veer_gene_2002}, there are 24,481 genes of 117 breast cancer patients in this dataset. For the 78 breast cancer patients with disease-free survival time, for each of the genes the $\log_{10}(Intensity)$ is given. We construct a regression model to select some important genes whose $\log_{10}(Intensity)$ is significantly related to disease-free patient survival. In this analysis, we regard the disease-free survival time as the response and use the $\log_{10}(Intensity)$ of genes as predictors.

It is assumed that only a few genes are associated with disease-free survival time. For very high-dimensional datasets, it is common to screen some variables first and then run variable selection based on the predictors we screened, such as in \cite{bai_variable_2021}. Thus we first compute the robust marginal correlations (using the GR estimator) of gene expressions with disease-free survival time and then screen the 100 genes that have the largest robust pairwise correlation coefficients as candidate predictors. 

\begin{figure}[!htp]
	\centering
	\includegraphics[width=14cm]{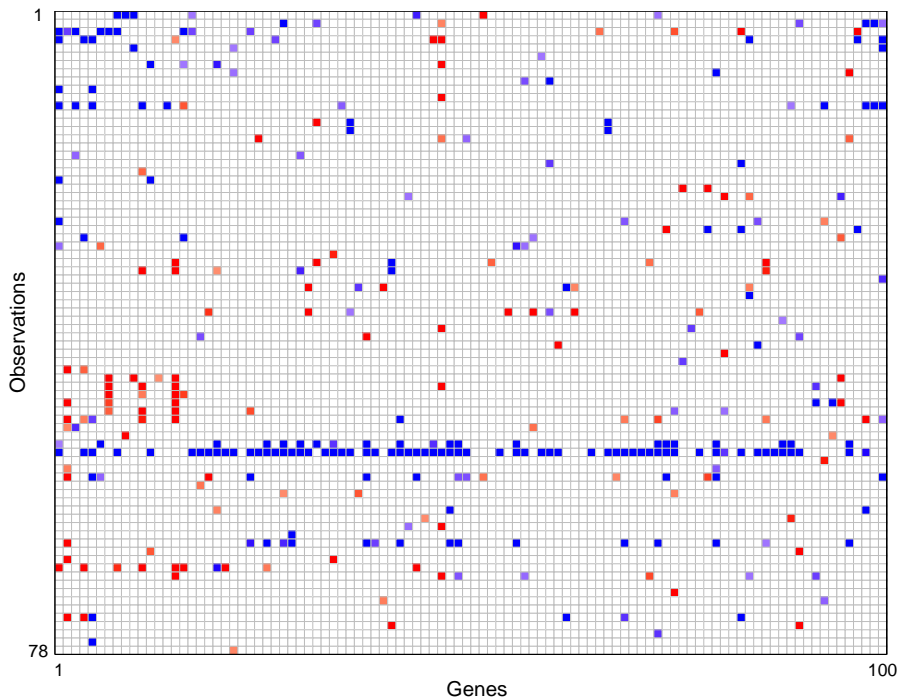}
	\caption{Outlier cell map for 100 screened variables on 78 patients from the breast cancer data. Most cells are blank, showing they are not detected as outliers. Cells are flagged as outlying if the observed value and predicted value differ too much. A red cell means the observed value is larger than the predicted value and a blue cell means the observed value is smaller than the predicted value significantly.}
	\label{breastcancer}
\end{figure}

Outlier detection results of the selected predictors based on DDC \cite{rousseeuw_detecting_2018} are shown in Figure \ref{breastcancer}. Most cells are blank, showing they are not detected as outliers. Cells are flagged as outlying (red and blue cells) if the observed value and predicted value differ too much. A red cell means the observed value is larger than the predicted value and a blue cell means the observed value is smaller than the predicted value significantly.

In this screened breast cancer dataset, some cells are detected as outlying. The average contamination rate for all 100 predictors is $4.50\%$. For most genes, the contamination rate is less than $5\%$. Some genes have a contamination rate of more than $10\%$. Looking at the outlier pattern in each row, that is for each patient, the contamination rates of most patients are below $5\%$. The $54$-th patient has the most detected outliers. Therefore, this row can be regarded as a rowwise outlier. 

Given the cellwise contamination, we first standardize all variables with the median and $Q_n$. Then a sparse regression model is fitted via GR-ALasso, MM-ALasso, ALasso, Lasso, sShootingS, sLTS and RLars, separately. In order to assess the performance of the selected models, we run a post-MM-estimator \cite{Yohai_1987} for each selected model and use Leave-One-Out Cross-Validation to measure the performance. The prediction accuracy of these methods is evaluated by the root of mean squared error (RMSPE), the mean absolute prediction error (MAPE) and the root of $90\%$ trimmed mean square prediction error (RTMSPE).
Model results are shown in Table \ref{cancerresult}.

\begin{table}[htbp]
	\centering
	\caption{Model sizes, RMSPEs, MAPEs and RTMSPEs of the compared methods for the Breast cancer dataset with post-MM-estimator and Leave-One-Out Cross-Validation. 
 }
	\resizebox{14cm}{!}{
		\begin{tabular}{rrrrrrrrr}
\toprule
 & GR-ALasso & MM-ALasso & ALasso & Lasso & sShootingS & sLTS & RLars\\ 
\midrule
Model size & 14 & 10 & 16 & 21 & 61 & 19 & 5\\ 
 MAPE & 20.55 & 19.54 & 21.72 & 26.94 & 56.13 & 20.71 & 21.49\\ 
 RMSPE & 24.66 & 27.86 & 25.80 & 36.52 & 72.04 & 26.49 & 28.49\\ 
 RTMSPE & 20.21 & 19.30 & 21.68 & 24.29 & 54.00 & 20.66 & 22.27\\ 
\bottomrule
\end{tabular}
	}%
	\label{cancerresult}%
\end{table}%

By examining the results presented in Table \ref{cancerresult}, it is apparent that the gene selection outcomes differ significantly across the various methods compared. Among all the methods evaluated, MM-ALasso demonstrates the most favourable performance by achieving the lowest prediction errors. 
On the other hand, Lasso, ALasso, RLars, and sLTS exhibit slightly inferior performance. The sShootingS method shows poor performance since it selects too many predictors, resulting in an overfitted model. GR-ALasso also exhibits good performance both in terms of prediction error and the number of features selected.

\section*{Funding}

Su's work was supported by the Chinese Scholarship Council (201906360181). Muller and Tarr were supported by the Australian Research Council (DP210100521). Tarr was supported in part by the AIR$@$innoHK program of the Innovation and Technology Commission of Hong Kong.

\bibliographystyle{tfnlm}
\bibliography{reference}

\newpage

\section*{A1. Proof of Theorem \ref{theorem1}}
\ps{Recall the definition $\bm{\rho} =\mathrm{vecl}(\bm{R})$ as provided in Section~\ref{theorem}, where $\mathrm{vecl}(\cdot)$ stands for a vector-type operator that arranges the elements in the lower triangle of a matrix in a column-wise manner. Let $\hat{\bm{R}}$ represent the estimated correlation matrix achieved through the Gaussian rank estimator. Consequently, we can establish an estimation denoted as $\hat{\bm{\rho}} =\mathrm{vecl}(\hat{\bm{R}}).$
We also define $\phi(\check{\bm{z}})$ as the probability density function (PDF) of random vector $\check{\bm{z}}$. The score function is given by
\[ s(\bm{\rho}) =\frac{\partial\log\phi(\check{\bm{z}})}{\partial\bm{\rho}},\]
which is followed by the Fisher information matrix, the variance of the score function,
\[\mathcal{I}(\bm{\rho}) =\text{Var}[s(\bm{\rho})].\]
}
When assumption A1 holds, according to \cite{amengual2022gaussian}, the asymptotic distribution of $\hat{\bm{R}}$ and the following $\bm{\rho}$ has the same asymptotic variance as the maximum likelihood estimator,
\[
\sqrt{n} (\hat{\bm\rho} -\bm{\rho})\overset{d}{\rightarrow} N(\bm 0,\mathcal{I}^{-1}(\bm{\rho})).
\]
\ps{Let $\check{\bm{\beta}}^{} =\bm{R}_{\bm{x}\bm{x}}^{-1}\bm{R}_{\bm{x}y}$ be the true standardized regression coefficient vector.} 
Using the delta method, the asymptotic distribution of $\hat{\check{\bm{\beta}}} =\hat{\bm R}_{\bm{x}\bm{x}}^{-1}\hat{\bm R}_{\bm{x}y}$ is,
\[
\sqrt{n}\left(\hat{\check{\bm{\beta}}} -\check{\bm{\beta}}^{}\right)\stackrel{d}{\rightarrow} N\left(\bm{0},\bm{\nabla}\check{\bm{\beta}}^{}\mathcal{I}^{-1}\left(\bm{\rho}\right)\bm{\nabla}\check{\bm{\beta}}^{ \top}\right),
\]
where $\bm{\nabla}\check{\bm{\beta}}^{} =\partial\check{\bm{\beta}}^{}/\partial\bm{\rho}^{\top}$ indicates the gradient of $\check{\bm{\beta}}^{}$ with respect to $\bm{\rho}$.
For each $z^j$, we obtain a consistent $Q_n$ estimator ${\hat\sigma}_{z^j}\overset{p}{\rightarrow}{\sigma}_{z^j}$. 
Note that,
$$
\bm{\beta}^{} =\bm{\Sigma}_{\bm{x}\bm{x}}^{-1}\bm{\Sigma}_{\bm{x}y} =\operatorname{Diag}^{-1}(\bm{\sigma}_{x})\check{\bm{\beta}}^{}\cdot\sigma_y,
$$ 
is the true parameter vector, where $\bm{\sigma}_{x} = (\sigma_{x^1},\ldots,\sigma_{x^p})^\top$, $x^j$ denotes the $j$-th variable of $\bm{x}$. By Slutsky's theorem, the asymptotic distribution of $\tilde{\bm\beta} =\hat{\bm{\Sigma}}_{\bm{x}\bm{x}}^{-1}\hat{\bm{\Sigma}}_{\bm{x}y}$ is,
\begin{equation}
\sqrt{n}\left(\tilde{\bm{\beta}}-\bm{\beta}^{}\right)\stackrel{d}{\rightarrow} N\left(\bm{0},\bm{\nabla}\bm{\beta}^{}\mathcal{I}^{-1}\left(\bm{\rho}\right)\bm{\nabla}\bm{\beta}^{ \top}\right),
\label{betaest}
\end{equation}
where 
\[
\bm{\nabla}\bm{\beta}^{} =\partial{\bm{\beta}^{}}/\partial{\bm{\rho}^{\top}} =\operatorname{Diag}^{-1}(\bm{\sigma}_{x})\bm{\nabla}\check{\bm{\beta}}^{}\cdot\sigma_y.
\]
To establish the asymptotic normality of the GR-ALasso estimator, let 
\ps{
\[ \bm{u} =\sqrt{n} ({\bm{b}} -\bm{\beta}^{}),\]
where $\bm{b}$ indicates possible estimates of $\bm{\beta}$.
}
Then we obtain,
\[{\bm{b}} ={\bm{\beta}}^{} +\frac{\bm u}{\sqrt{n}},\] 
and 
\[
 \psi(\bm{u}) = n\left\|\bm{v}-\bm{W}\left(\bm{\beta}^{}+\frac{\bm u}{\sqrt{n}}\right)\right\|_2^{2}+\lambda\sum_{j=1}^{p}{\omega}_{j}\left|\beta_{j}^{}+\frac{u_{j}}{\sqrt{n}}\right|.
\]
Let 
\[
\hat{\bm{u}}=\mathop{\operatorname{argmin}}\limits_{\bm{u}} \psi(\bm{u}),
\]
then we obtain,
$$\hat{\bm{\beta}}=\bm{\beta}^{}+\frac{\hat{\bm{u}}}{\sqrt{n}},$$ 
and
$$\hat{\bm{u}}=\sqrt{n}\left(\hat{\bm{\beta}}-\bm{\beta}^{}\right).$$ 
Define $V(\bm{u}) = \psi(\bm{u})- \psi(\bm 0)$. Then,

\[
\begin{aligned}
V(\bm{u})\ \equiv\ &\bm{u}^{\top}\left(\bm{W}^{\top}\bm{W}\right)\bm{u}-2{{\sqrt{n}}(\bm v -\bm{W}\bm{\beta}^{})^{\top}\bm{W}}\bm{u}\\
& +\frac{\lambda}{\sqrt{n}}\sum_{j=1}^{p}{\omega}_{j}\sqrt{n}\left(\left|\beta_{j}^{}+\frac{u_{j}}{\sqrt{n}}\right|-\left|\beta_{j}^{}\right|\right)\\
\triangleq & V_{1}(\bm{u}) \ps{- 2V_{2}(\bm{u})} + V_{3}(\bm{u}).
\end{aligned}
\]
For $V_{1}(\bm{u})$, it is straightforward to see that $\bm{W}^{\top}\bm{W} = \ps{\hat{\bm{\Sigma}}_{\bm{x}\bm{x}}}$ and $\hat{\bm{\Sigma}}_{\bm{x}\bm{x}}\overset{p}{\rightarrow}\bm{\Sigma}_{\bm{x}\bm{x}}$. 
For $ V_{2}(\bm{u})$, we see that,
\[
\begin{aligned}
V_{2}(\bm{u}) = &\sqrt{n}\left (\hat{\bm{\Sigma}}_{\bm{x}y} -\hat{\bm{\Sigma}}_{\bm{x}\bm{x}}\bm{\beta}^{}\right )^\top\bm{u}\\
= &\sqrt{n}\bm{u}^\top\hat{\bm{\Sigma}}_{\bm{x}\bm{x}}\left (\hat{\bm{\Sigma}}_{\bm{x}\bm{x}} ^{-1}\hat{\bm{\Sigma}}_{\bm{x}y} -\bm{\beta}^{}\right ).\\
\end{aligned}
\]
Then, by Slutsky’s theorem, we obtain,
\[
\sqrt{n}\hat{\bm\Sigma}_{\bm{x}\bm{x}}\left (\hat{\bm{\Sigma}}_{\bm{x}\bm{x}} ^{-1}\hat{\bm{\Sigma}}_{\bm{x}y} -\bm{\beta}^{}\right ) \ps{\overset{d}{\rightarrow}}\bm{D} \ps{\overset{d}{=}} N\left(\bm{0},\bm{\Sigma}_{\bm{x}\bm{x}}\bm{\nabla}\bm{\beta}^{}\mathcal{I}^{-1}\left(\bm{\rho}\right)\bm{\nabla}\bm{\beta}^{ \top}\bm{\Sigma}_{\bm{x}\bm{x}}\right).
\]
Now let us consider the behaviour of the third term $V_{3}(\bm{u})$. If $\beta_j\neq 0$, we have,
$$
\omega_j = |\tilde\beta_j|^{-1}\overset{p}{\rightarrow } |\beta_j^{}|^{-1},
$$
and 
$$
\sqrt{n}\left(\left|\beta_{j}^{}+\frac{u_{j}}{\sqrt{n}}\right|-\left|\beta_{j}^{}\right|\right) = u_{j}\operatorname{sign}\left(\beta_{j}^{}\right).
$$
According to Slutsky's theorem and the assumption that ${\lambda}/{\sqrt{n}} \rightarrow 0$ , we have 
$$
\frac{\lambda}{\sqrt{n}}{\omega}_{j}\sqrt{n}\left(\left|\beta_{j}+\frac{u_{j}}{\sqrt{n}}\right|-\left|\beta_{j}\right|\right)\overset{p}{\rightarrow} 0.
$$
If $\beta_j^{} = 0$, then we have,
$$
\sqrt{n}\left(\left|\beta_{j}^{}+\frac{u_{j}}{\sqrt{n}}\right|-\left|\beta_{j}^{}\right|\right) = |u_{j}|,
$$ and 
$$
\frac{\lambda}{\sqrt{n}}{\omega}_{j}={\lambda}\left(\left|\sqrt{n} \ps{\tilde{\beta}_{j}}\right|\right)^{-1},
$$
where $\sqrt{n} \ps{\tilde{\beta_j}} = O_p(1)$. In this case, $V_{3}(\bm{u})\rightarrow\infty$ since $\lambda\rightarrow\infty$. Combining these conclusions together, we have \ps{$V(\bm{u})\overset{d}{\rightarrow} K(\bm{u})$} for every $\bm{u}$,
\[
 K(\bm{u})=\left\{\begin{array}{ll}
\bm{u}_{\mathcal{A}}^{\top}\bm{\Sigma}_{\mathcal{A}}\bm{u}_{\mathcal{A}}-2\bm{u}_{\mathcal{A}}^{\top}\bm{D}_{\mathcal{A}} &\text{if } u_{j}=0\ \forall j\notin\mathcal{A}\\
\infty &\text{otherwise, }
\end{array}\right.
\]
where 
$\bm{u}_{\mathcal{A}}$, $\bm{\Sigma}_{\mathcal{A}}$ and $\bm{D}_{\mathcal{A}}$ indicate $\bm{u}$, $\bm{\Sigma}$ and $\bm{D}$ with components only related to variables in $\mathcal{A}$.
Note that \ps{$K(\bm{u})$} is convex and the unique optimization of \ps{$K(\bm{u})$} is,
$$
((\bm{\Sigma}_{\mathcal{A}}^{-1}\bm{D}_{\mathcal{A}})^\top,\bm{0}^\top)^\top.
$$
According to the epi-convergence results of \cite{geyer1994asymptotics} and \cite{fu2000asymptotics}, we obtain,
$$
\hat{\bm{u}}_{\mathcal{A}}\overset{d}{\rightarrow}\bm{\Sigma}_{\mathcal{A}}^{-1}\bm{D}_{\mathcal{A}},
$$
and 
$$
\hat{\bm{u}}_{\mathcal{A}^c}\overset{d}{\rightarrow}\bm{0},
$$
 \ps{where ${\mathcal{A}^c}$ is the complement set of $\mathcal{A}$, $\hat{\bm{u}}_{\mathcal{A}}$ represents a subvector of $\hat{\bm{u}}_{\mathcal{A}}$ containing only the components related to variables in $\mathcal{A}$, while $\hat{\bm{u}}_{\mathcal{A}^c}$ represents a subvector of $\hat{\bm{u}}$ containing only the components related to variables in the complement set of $\mathcal{A}$.}
Thus we prove the asymptotic normality of the GR-ALasso estimator.

The proof for consistency follows from the asymptotic normality we obtained above. For all $j\in\mathcal{A}$, we have $P\left(j\in\hat{\mathcal{A}}\right)\rightarrow 1$. Then, for $\forall j^{\prime}\notin\mathcal{A}$, we should prove that $P\left(j^\prime\in\hat{\mathcal{A}}\right)\rightarrow 0$. 
Considering the event $j^{\prime}\in\hat{\mathcal{A}}$, we denote $x_{j^{\prime}}$ as the corresponding variable. By the KKT conditions, we have,
$$
\bm{W}_{{j^\prime}}^\top\bm{v} -\bm{W}_{{j^\prime}}^\top\bm{W}\hat{\bm{\beta}} =\hat\Sigma_{x_{j^\prime}y} -\hat{\bm\Sigma}_{x_{j^\prime}\bm{x}}^\top\hat{\bm{\beta}} =\lambda\omega_{j\prime}/2n, 
$$ 
 \ps{where $\bm{W}_{{j^\prime}}$ indicates the $j^\prime$-th column of $\bm{W}$ corresponding to variable $x_{j^\prime}$, $\hat\Sigma_{x_{j^\prime}y}$ indicates the estimated covariance between $x_{j^\prime}$ and $y$, $\hat{\bm\Sigma}_{x_{j^\prime}\bm{x}}$ indicates an estimated covariance between $x_{j^\prime}$ and $\bm{x}$, $\omega_{j\prime}$ indicates the adaptive weight of the $j^\prime$-th variable.}
Then we obtain,
$$
\lambda\hat{w}_{j^{\prime}} /\sqrt{n}=\lambda/{\left|\sqrt{n}\tilde{\beta}_{j^{\prime}}\right|}\overset{p}{\rightarrow}\infty,
$$ 
as $\lambda\rightarrow\infty$. Meanwhile, we see that,
 \ps{
\[
\begin{aligned}
\sqrt{n}(\hat\Sigma_{x_{j^\prime}y} -\hat{\bm\Sigma}_{x_{j^\prime}\bm{x}}^\top\hat{\bm{\beta}}) &=\sqrt{n}\hat{\bm\Sigma}_{x_{j^\prime}\bm{x}}\left(\hat{\bm{\Sigma}}_{\bm{x}\bm{x}} ^{-1}\hat{\bm{\Sigma}}_{\bm{x}y} -\hat{\bm{\beta}}\right)\\
&=\sqrt{n}\hat{\bm\Sigma}_{x_{j^\prime}\bm{x}}\left((\hat{\bm{\Sigma}}_{\bm{x}\bm{x}} ^{-1}\hat{\bm{\Sigma}}_{\bm{x}y} -\bm{\beta}^{}) - (\hat{\bm{\beta}}-\bm{\beta}^{})\right).
\end{aligned}
\]
Following the conclusions we obtained above, we establish that both $\sqrt{n}\hat{\bm\Sigma}_{x_{j^\prime}\bm{x}}\left(\hat{\bm{\Sigma}}_{\bm{x}\bm{x}} ^{-1}\hat{\bm{\Sigma}}_{\bm{x}y} -\bm{\beta}^{}\right)$ and $\sqrt{n}\hat{\bm\Sigma}_{x_{j^\prime}\bm{x}}\left(\hat{\bm{\beta}}-\bm{\beta}^{}\right)$ converge to normal distributions.
Hence $\sqrt{n}(\hat\Sigma_{x_{j^\prime}y} -\hat{\bm\Sigma}_{x_{j^\prime}\bm{x}}^\top\hat{\bm{\beta}})$ converge to a normal distribution. 
}
Thus, as in \cite{zou_adaptive_2006}, we have 
\[
\begin{aligned}
P(j^\prime\in\hat{\mathcal{A}}) &\leq P\left(\hat\Sigma_{x_{j^\prime}y} -\hat{\bm\Sigma}_{x_{j^\prime}\bm{x}}^\top\hat{\bm{\beta}} =\frac{\lambda\omega_{j^\prime}}{2n}\right)\\
& = P\left(\sqrt{n}\left(\hat\Sigma_{x_{j^\prime}y} -\hat{\bm\Sigma}_{x_{j^\prime}\bm{x}}^\top\hat{\bm{\beta}}\right) =\frac{\lambda\omega_{j^\prime}}{2\sqrt{n}}\right)\\
&\rightarrow 0.
\end{aligned}
\]
Therefore the consistency of the GR-ALasso estimator is proved.

\newpage
\section*{A2. Supplementary Figures}

\begin{figure}[!htp]
	\centering
	\includegraphics[width=14cm]{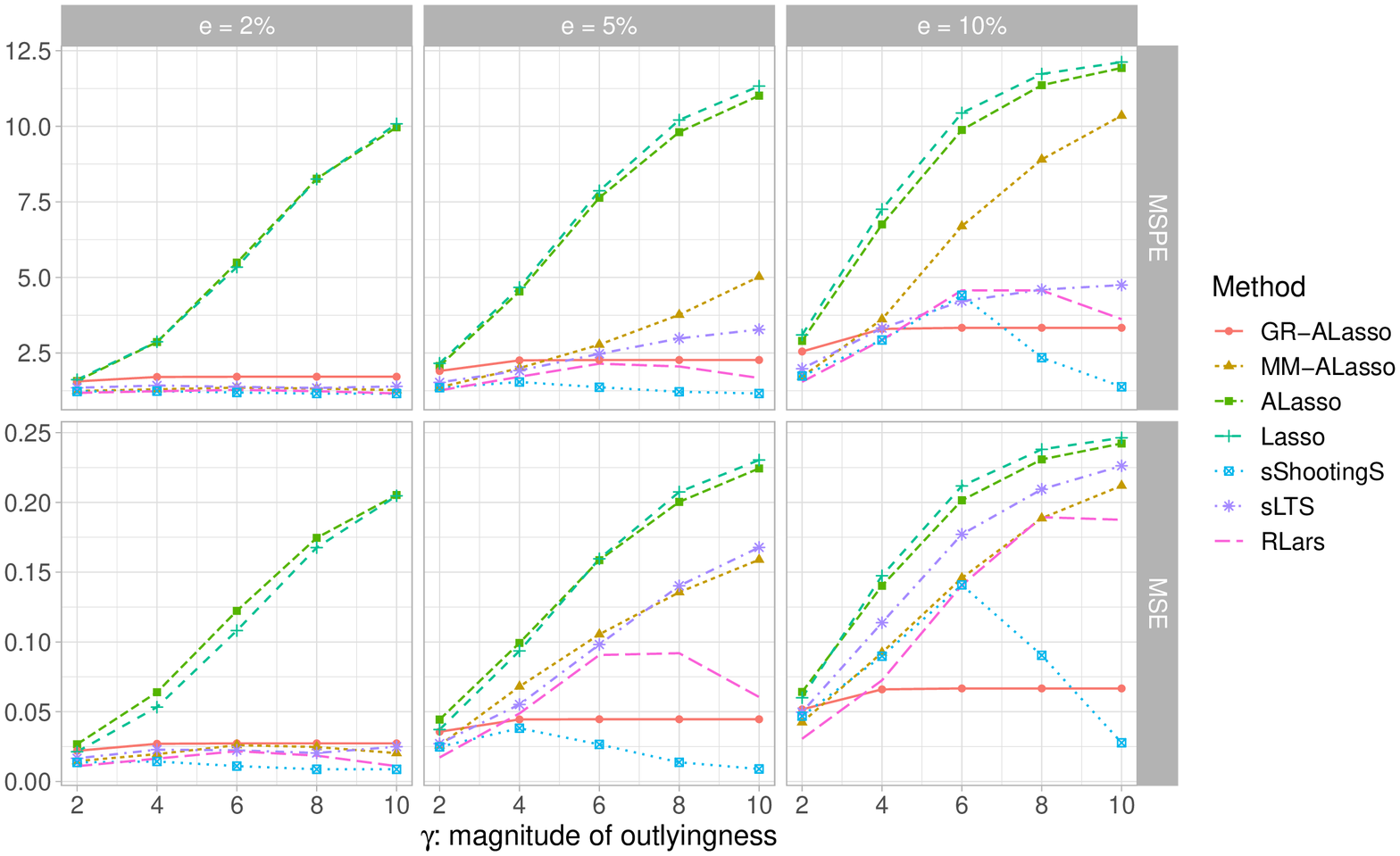}
	\caption{The averages of MSPEs and MSEs over 200 simulation runs for various contamination rates when $p = 20$. Each column represents a contamination rate. The horizontal axis represents the magnitude of outlyingness and the vertical axis gives the MSPEs and MSEs.}
\end{figure}

\begin{figure}[!htp]
	\centering
	\includegraphics[width=14cm]{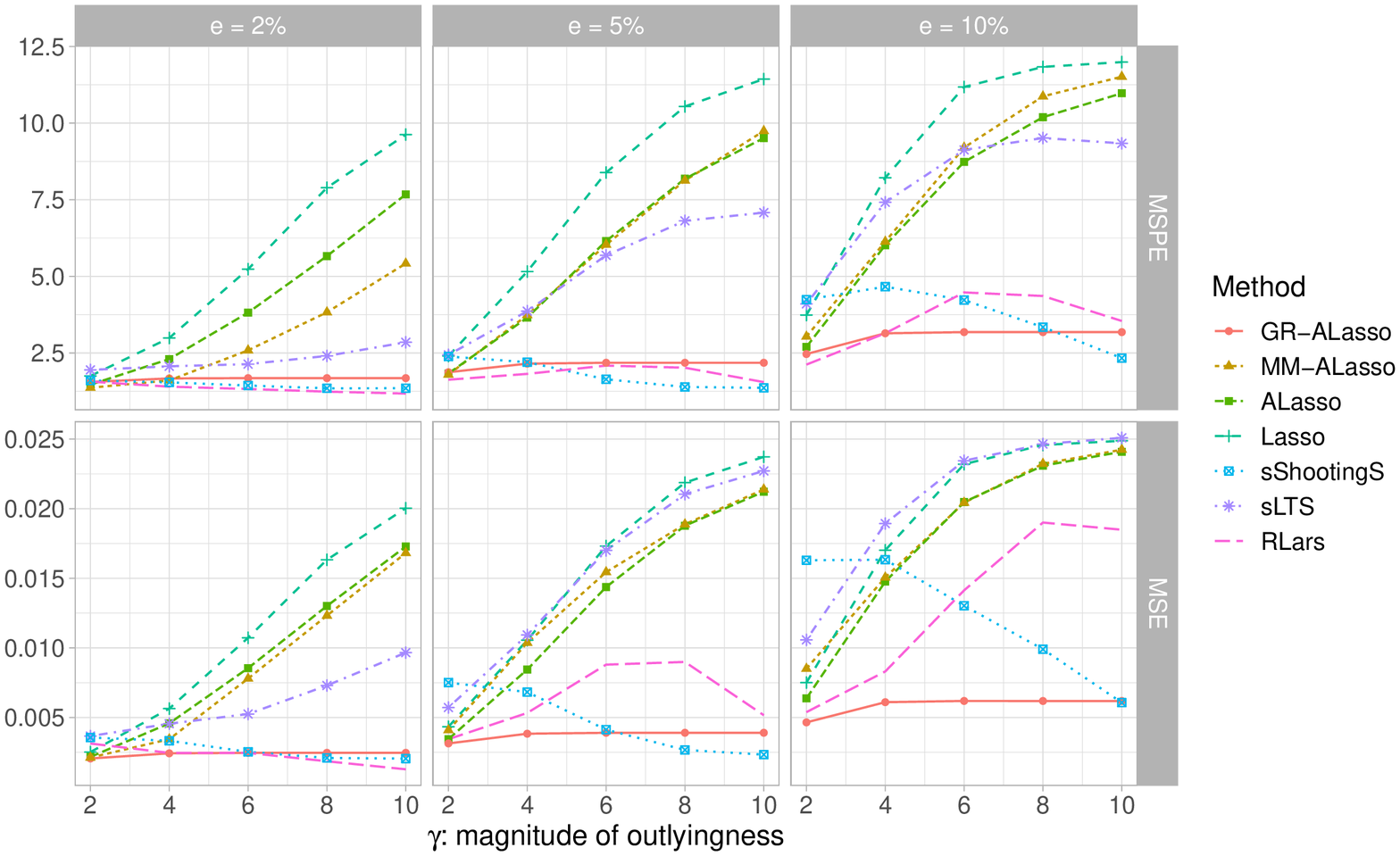}
	\caption{The averages of MSPEs and MSEs over 200 simulation runs for various contamination rates when $p = 200$. Each column represents a contamination rate. The horizontal axis represents the magnitude of outlyingness and the vertical axis gives the MSPEs and MSEs.}
\end{figure}

\appendix

\end{document}